# Augmenting Heritage: An Open-Source Multiplatform AR Application

This preprint has not undergone peer review or any post-submission corrections and is being reviewed at an appropriate journal.

Corrie Green[1]
Robert Gordon University
c.green1@rgu.ac.uk

June 20th, 2023

ABSTRACT

AI NeRF algorithms, capable of cloud processing, have significantly reduced hardware requirements and processing efficiency in photogrammetry pipelines. This accessibility has unlocked the potential for museums, charities, and cultural heritage sites worldwide to leverage mobile devices for artifact scanning and processing. However, the adoption of augmented reality platforms often necessitates the installation of proprietary applications on users' mobile devices, which adds complexity to development and limits global availability. This paper presents a case study that demonstrates a cost-effective pipeline for visualizing scanned museum artifacts using mobile augmented reality, leveraging an open-source embedded solution on a website.

## 1 Introduction

The loss of the additional dimension when visualizing 3D scanned objects on a 2D screen can lead to a reduction of user insight and engagement when compared to physically interacting or viewing museum artifacts. Many museums hold collections that are not available for interaction due to being behind a display case or aren't available due to a lack of exhibition space.

This paper presents a cost-effective pipeline for adopting 3D photogrammetry scans into a browser-based solution allowing for display on both mobile and desktop devices using Google's 3D Model viewer framework supporting Augmented Reality (AR) [1]. Due to the low-cost options available of high-fidelity scanning technology today, there has been a mass adoption of digital twinning of cultural heritage sites and museum artifacts for preservation [2], allowing future generations to experience historically rich sites that may have since been damaged or lost.

[1] https://orcid.org/0000-0003-0404-3668.

By 3D scanning artifacts using photogrammetry or lidar-based approaches, we can accurately reproduce a digital twin for an artifact or historical site for use in visualization and interactive applications. Although scanning technologies have improved, providing an engaging and accessible presentation environment for scanned models has been lagging.

Mobile augmented reality allows users with smartphone devices to augment information in their real-world environment. In addition to overlaying information such as text, we can present access to the third dimension for artifacts that are traditionally behind glass protection in museum exhibits allowing visitors to explore all its perspectives.

Without an accessible and always available database of models, scanned objects may not be fully exploited by members of the public and research community. In addition, due to the closure of museums around the world during global pandemics, artifacts have been preserved without physical access. A generous interface would support in representing the richness of cultural heritage collections, allowing for visualization for visitors around the world at any time to explore and enrich interpretation between associated collections [3].

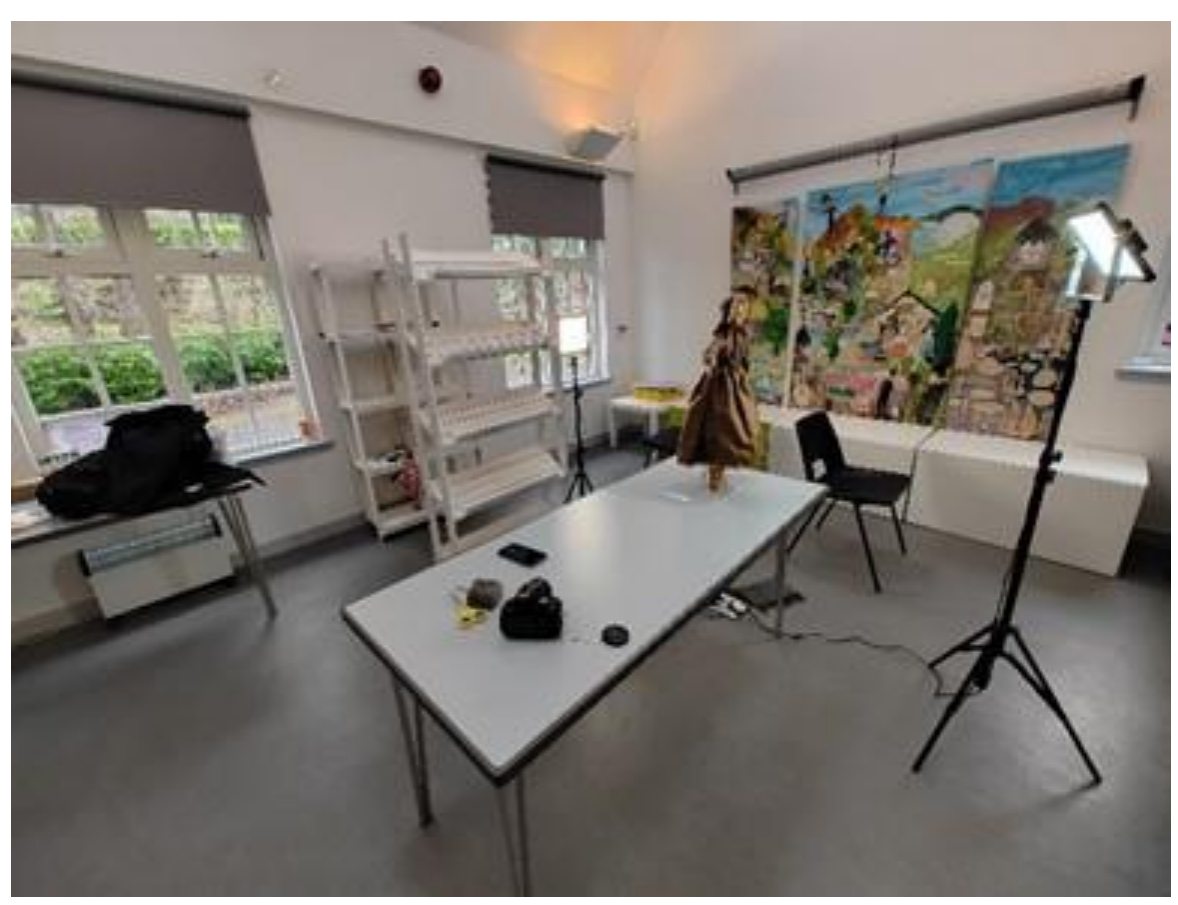

Figure 1: Photo of the Scanning environment made available at the Museum of Childhood Dingwall

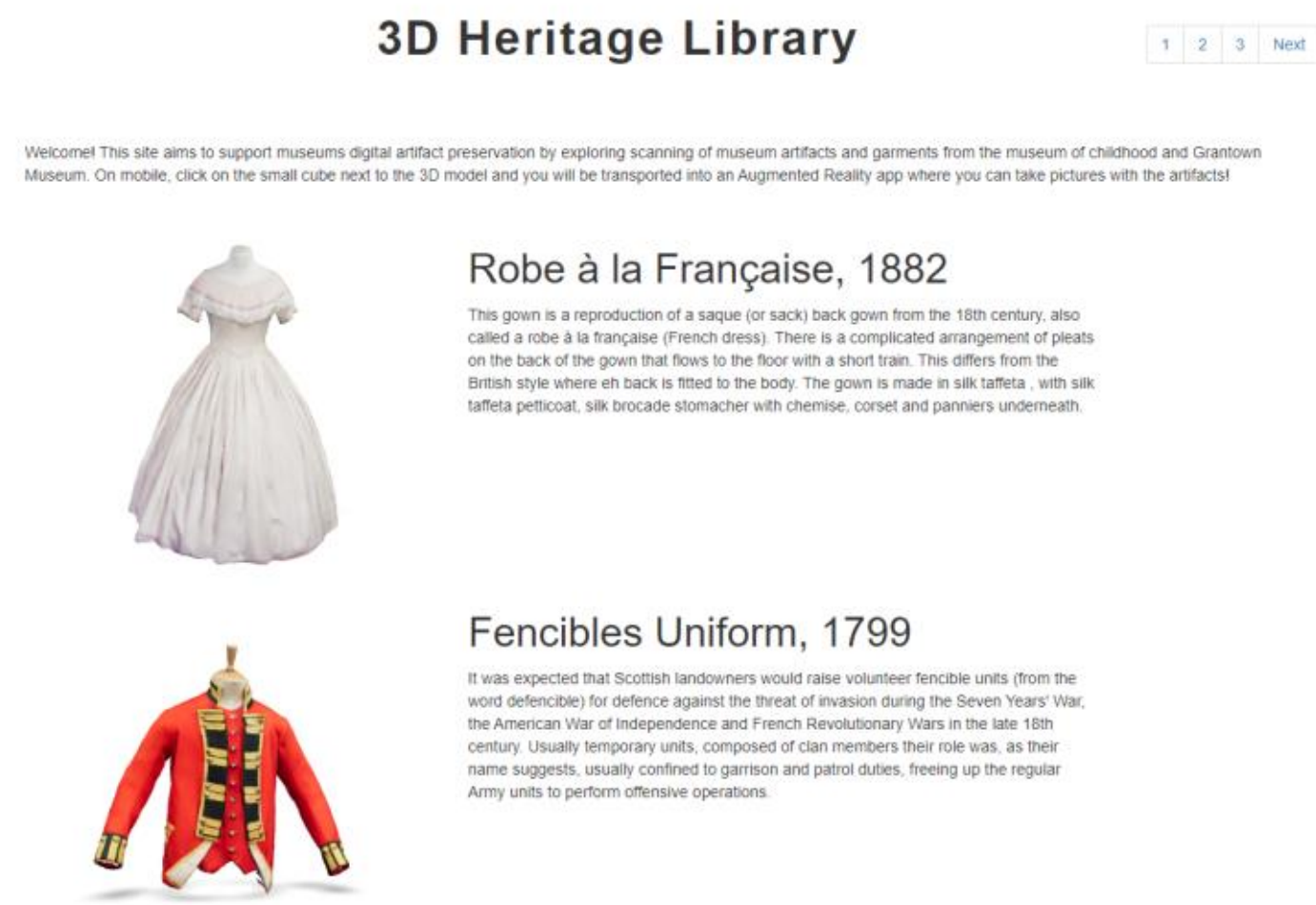

Figure 2: Screenshot of the developed website being rendered on a desktop browser

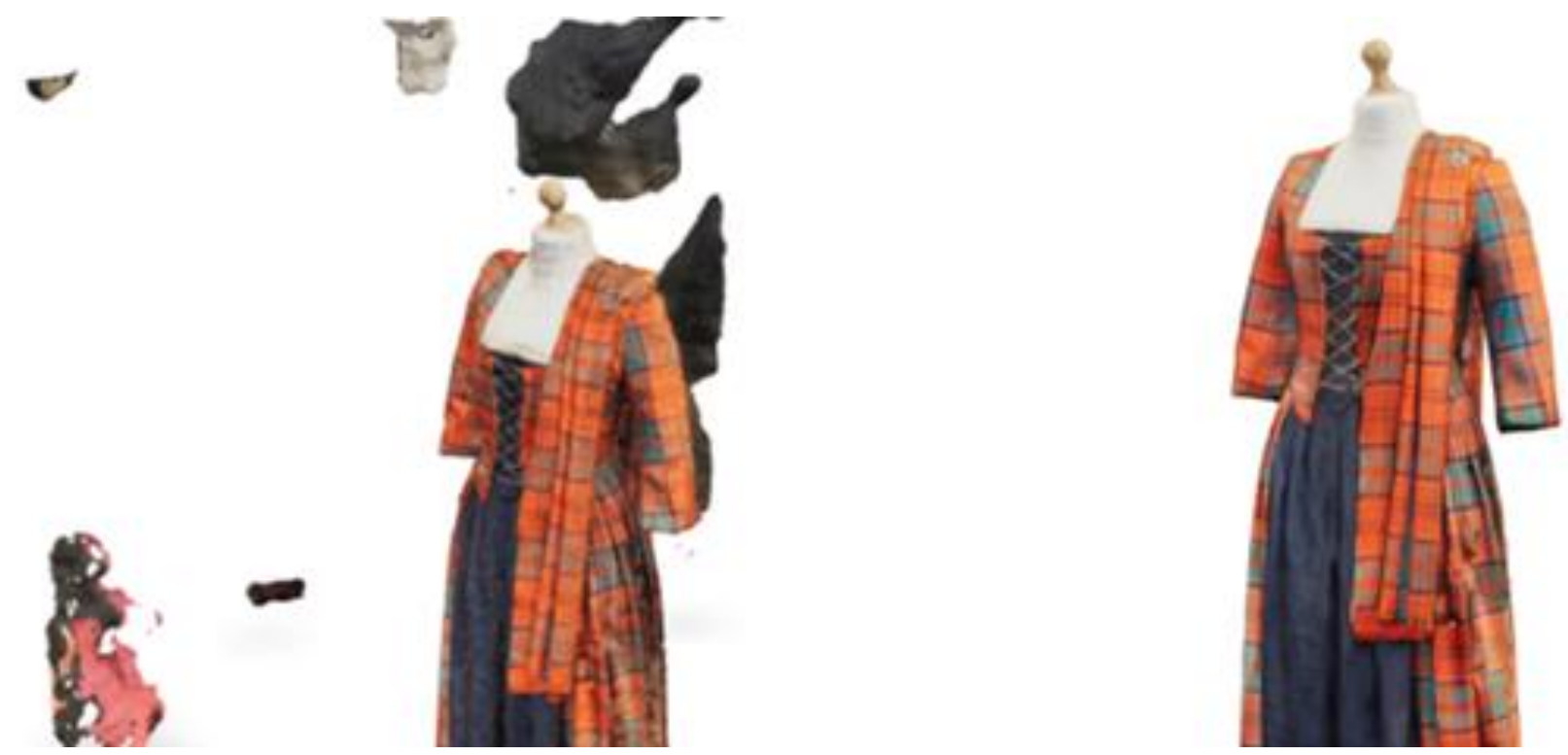

Figure 3: Screenshot demonstrated the unprocessed 3D scan without data cleanup of a Scottish dress with a comparison of the cleaned scan on the right

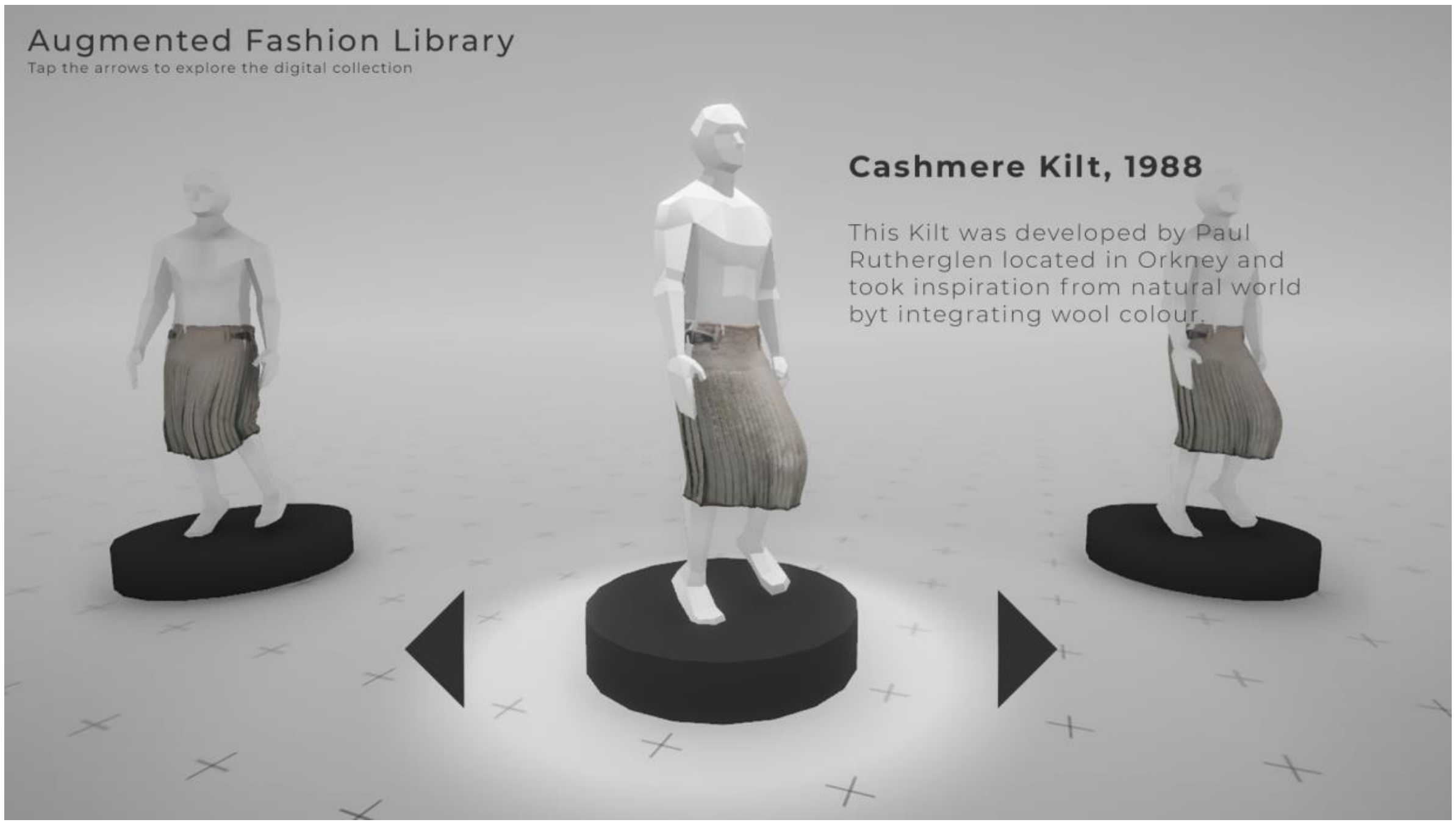

Figure 6 Demonstrates how a scanned kilts mesh renderer can be overlayed onto a humanoid character with animations for use in game engines.

Link to the project repository and website: https://github.com/corriedotdev/3D-Museum-Library